\documentclass[pra,showpacs,amsmath,amssymb,twocolumn, 10pt]{revtex4}
\usepackage{enumerate}
\usepackage{color}
\usepackage{amsthm}
\usepackage{dcolumn}
\usepackage{bm}
\usepackage{graphicx}

\begin{document}

\newtheorem{corollary}{Corollary}
\newtheorem{definition}{Definition}
\newtheorem{example}{Example}
\newtheorem{lemma}{Lemma}
\newtheorem{proposition}{Proposition}
\newtheorem{theorem}{Theorem}
\newtheorem{fact}{Fact}
\newtheorem{property}{Property}
\newcommand{\bra}[1]{\langle #1|}
\newcommand{\ket}[1]{|#1\rangle}
\newcommand{\braket}[3]{\langle #1|#2|#3\rangle}
\newcommand{\ip}[2]{\langle #1|#2\rangle}
\newcommand{\op}[2]{|#1\rangle \langle #2|}

\newcommand{\tr}{{\rm tr}}
\newcommand{\supp}{{\it supp}}
\newcommand{\sch}{{\it Sch}}

\newcommand{\Span}{\mathrm{span}}
\newcommand {\E } {{\mathcal{E}}}
\newcommand {\G } {{\mathcal{G}}}
\newcommand{\In}{\mathrm{in}}
\newcommand{\Out}{\mathrm{out}}
\newcommand{\local}{\mathrm{local}}
\newcommand {\F } {{\mathcal{F}}}
\newcommand {\diag } {{\rm diag}}
\renewcommand{\b}{\mathcal{B}}
\newcommand{\h}{\mathcal{H}}
\renewcommand{\Re}{\mathrm{Re}}
\renewcommand{\Im}{\mathrm{Im}}
\newcommand{\Z}{\sigma_z}
\newcommand{\Clocal}{C^{(0)}_{\local}}
\newcommand{\Sp}[1][p]{S_{\min}^{(#1)}}

\title{Super-Activation of Zero-Error Capacity of Noisy Quantum Channels}
\author{Runyao Duan}
\email{Runyao.Duan@uts.edu.au} \affiliation{State Key Laboratory of
Intelligent Technology and Systems,Tsinghua National Laboratory for
Information Science and Technology, Department of Computer Science
and Technology, Tsinghua University, Beijing 100084, China and
\\
Center for Quantum Computation and Intelligent Systems (QCIS),
Faculty of Engineering and Information Technology, University of
Technology, Sydney, NSW 2007, Australia}

\begin{abstract}
We study various super-activation effects in the following
zero-error communication scenario: One sender wants to send
classical or quantum information through a noisy quantum channel to
one receiver with zero probability of error. First we show that
there are quantum channels of which a single use is not able to
transmit classical information perfectly yet two uses can. This is
achieved by employing entangled input states between different uses
of the given channel and thus cannot happen for classical channels.
Second we exhibit a class of quantum channel with vanishing
zero-error classical capacity such that when a noiseless qubit
channel or one ebit shared entanglement are available, it can be
used to transmit $\log_2 d$ noiseless qubits, where $2d$ is the
dimension of input state space. Third we further construct quantum
channels with vanishing zero-error classical capacity when assisted
with classical feedback can be used to transmit both classical and
quantum information perfectly. These striking findings not only
indicate both the zero-error quantum and classical capacities of
quantum channels satisfy a strong super-additivity beyond any
classical channels, but also highlight the activation power of
auxiliary physical resources in zero-error communication.
\end{abstract}


\maketitle
\section{Introduction}
The notion of zero-error capacity was introduced by Shannon in 1956
to characterize the ability of noisy channels to transmit classical
information with zero probability of error \cite{SHA56}. Since
Shannon's seminal work, the study of this notion and the related
topics has grown into a vast field called \textit{zero-error
information theory} \cite{KO98}. The main motivation is partly due
to the following facts: (1) In many real-world critical applications
no errors can be tolerated; (2) In practice, the communication
channel can only be available for a finite number of times; (3) Deep
connections to other research fields such as graph theory and
communication complexity theory have been established \cite{LOV79,
HAE79,ALO98,CRST06}. These works indicate that unlike the ordinary
capacity, computing the zero-error capacity of classical channels is
essentially a combinatorial optimization problem about graphs, and
is extremely difficult even for very simple graphs. Despite the fact
that numerous interesting and important results have been reported
(see \cite{KO98} for an excellent review), the theory of zero-error
capacity is still far from complete even for classical channels.

The generalization of zero-error capacity to quantum channels is
somewhat straightforward but nontrivial as the input states of the
channel may be entangled between different uses, and the information
transmitted may be classical or quantum. At least two notions of
zero-error capacity of quantum channels exist: one is the zero-error
classical capacity, the least upper bound of the rates at which one
can send classical information perfectly through a noisy quantum
channel, denote $C^{(0)}$. If replacing classical information with
quantum information in the definition of $C^{(0)}$, we have another
notion $Q^{(0)}$, the zero-error quantum capacity. A careful study
of these generalizations will not only help us to exploit new
features of quantum information, but also be useful in building
highly reliable communication networks. The notion of $Q^{(0)}$ has
been extensively investigated in the context of quantum-error
correction. In this paper we mainly focus on $C^{(0)}$ of which
little was known. A few preliminary works have been done towards to
a better understanding of the zero-error classical capacity of
quantum channels. In particular, some basic properties of $C^{(0)}$
of quantum channels were observed in \cite{MA05}. Later, it was
shown that the zero-error classical capacity for quantum channels is
in general also extremely difficult to compute~\cite{BS07}. However,
in these works the only allowable input states for channels were
restricted to be product states and entangled uses of the channel
were prohibited. Consequently, many of the properties of this notion
is similar to the classical case and it was not clear what kind of
role the additional quantum resources such as entanglement will play
in zero-error communication.

In a recent work it was demonstrated that the zero-error classical
capacity of quantum channels behaves dramatically different from the
corresponding classical capacity \cite{DS08}. More precisely, it was
shown that in the so-called multi-user communication scenario, there
is noisy quantum channel of which one use cannot transmit any
classical information perfectly yet two uses can. To achieve this,
one needs to encode the classical message using entangled states as
input and thus to make two uses of the channel entangled. This is a
purely quantum effect that cannot happen for any classical channels.
Furthermore, it cannot be observed under the assumptions of Refs.
\cite{MA05,BS07} where only product input states between different
uses are allowed. One drawback of the channel constructed in
\cite{DS08} is that we have at least two senders or two receivers
and require the senders or the receivers to perform local operations
and classical communication (LOCC) only. This LOCC restriction is a
reasonable assumption in practice as it captures the fact that the
quantum communication among the senders or the receivers would be
relatively expensive. If this local requirement is removed, one use
of these channels are able to transmit classical information
perfectly. Thus a major open problem left is to ask whether there is
quantum channel with only one sender and one receiver enjoying the
same property.

\section{Main Results}

\begin{figure}[ht]
  \centering
  \includegraphics[scale=0.5]{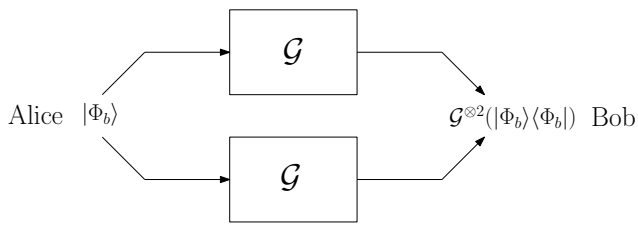}
  \caption{
  $\E$ is a noisy quantum channel from Alice to Bob. With one use of $\G$,
  Alice cannot transmit classical information to Bob perfectly.
  Interestingly, by using $\G$ twice, Alice can transmit a classical bit ``$b$" perfectly to
  Bob. To do so, Alice carefully encodes the bit ``$b$" into a bipartite entangled state $\ket{\Phi_b}$
and applies $\G$ twice. By decoding the output state $\G^{\otimes
2}(\op{\Phi_b}{\Phi_b})$, Bob can perfectly recover the bit ``$b$".}
  \label{pic1}
\end{figure}

The purpose of this paper is to further develop the theory of
zero-error capacity for quantum channels. Our first main result
(Theorem \ref{th1}) is an affirmative answer to the above open
problem. More precisely, we show by an explicit construction that
there does exist quantum channel $\G$ with one sender and one
receiver such that one use of $\G$ cannot transmit classical
information perfectly while two uses of $\G$ can transmit at least
one bit without any error. Fig. \ref{pic1} demonstrates our
construction. In our construction we don't construct $\G$ directly.
Instead, we construct two quantum channels $\E$ and $\F$ such that
both of them cannot transmit classical information perfectly by a
single use while can transmit at least one bit if employed jointly.
This confirms the usefulness of entangled input for perfect
transmission of classical information.

Similar to the previous work \cite{DS08}, our main tool is the
notion of unextendible bases (or equivalently, completely entangled
subspaces) \cite{PAR04,HLW06,DFJY07,CMW07, WS07}. The key ingredient
in our construction is to partition a bipartite Hilbert space into
two orthogonal subspaces which are both completely entangled, or
equivalently, unextendible. This kind of partitions has been found
before \cite{DS08, DXY07, CHL+08} and has been demonstrated very
useful in quantum information theory\cite{CHL+08,DXY07,DS08,WS07}.
However, all these previous partitions are not sufficient for our
purpose. Additional requirements make the construction rather
difficult and tricky.

Our second main result (Theorem \ref{th3}) is to show that both the
zero-error quantum and classical capacities of noisy quantum
channels are strongly super-additive. This is achieved by
introducing a class of special quantum channels which can be treated
as the generalizations of retro-correctible channels \cite{BDSS06}.
It was known that the zero-error capacity of classical channels are
super-additive in the following sense \cite{HAE79,ALO98}:There are
$\mathcal{N}_0$ and $\mathcal{N}_1$ such that
$C^{(0)}(\mathcal{N}_0\otimes
\mathcal{N}_1)>C^{(0)}(\mathcal{N}_0)+C^{(0)}(\mathcal{N}_1)$.  This
is very different from the ordinary classical capacity of classical
channels, which is always additive. However, any classical channels
$\mathcal{N}_0$ and $\mathcal{N}_1$ satisfying the super-additivity
must have the ability to transmit classical information perfectly,
that is $C^{(0)}(\mathcal{N}_0)>0$ and $C^{(0)}(\mathcal{N}_1)>0$.
It remains unknown whether the above super-additivity still holds if
one of the quantum channels are with vanishing zero-error capacity.
Here we show that for quantum channels such type of stronger
super-additivity can exist. Actually, we show that there are quantum
channels $\E$ and $\F$ such that $C^{(0)}(\E)=0$,
$Q^{(0)}(\F)=C^{(0)}(\F)=1$, but $Q^{(0)}(\E\otimes\F)=\log_2 d>
C^{(0)}(\E)+C^{(0)}(\F)=1$, where $2d$ is dimension of the input
state space of $\E$. The channel $\F$ can be chosen as a noiseless
qubit channel. If we are only concerned with zero-error classical
capacity, then $\E$ can be made entanglement-breaking (Theorem
\ref{th2}). Furthermore, if a $2\otimes 2$ maximally entangled state
is shared between the sender and the receiver or allowing two-way
classical communication that is independent from the message sending
from the main protocol, one use of $\E$ can be used to send $\log_2
d$ noiseless qubits. This type of $\E$ has the following weird
property: It is not able to communicate any classical information
perfectly; however, with a small amount of auxiliary resources (such
as one noiseless qubit channel, or one ebit, or two-way classical
communication independent from the messages sending through main
protocol), the channel acts as a noiseless quantum channel with
large perfect quantum capacity (achieving zero-error quantum
capacity $\log_2 d$). Intuitively, the hiding zero-error
communication ability of channel can be activated by these auxiliary
resources.

Our last main result is to study the role of classical feedback in
zero-error communication. As pointed out by Shannon, for classical
channels, the classical feedback cannot increase the ordinary
channel capacity but may increase the zero-error capacity
\cite{SHA56}. However, a necessary condition for such a feedback
improvement is that the channel should be able to communicate
classical information perfectly, i.e., with non-vanishing zero-error
capacity. It is of great interest to ask that whether this
requirement can be removed for quantum channels. Surprisingly, this
answer is yes. Specifically, we construct a quantum channel with a
two-dimensional input state space and vanishing zero-error classical
capacity such that when assisted with classical feedback enables
perfect transmission of classical and quantum information (Theorem
\ref{cfb}). In other words, the zero-error capacity of quantum
channels can be activated from $0$ to positive by classical
feedback. This remarkable phenomenon, demonstrates that the
zero-error communication ability of a quantum channel may be
recovered when assisted with classical feedback.

We notice that very recently several important super-activation
effects about different type of capacities of quantum channels,
namely quantum capacity, classical capacity, and the private
capacity, were discovered \cite{SY08, HAS09, LWZG09, SS09}. Clearly,
these results are incomparable to ours due to the special zero-error
transmission requirement.

\section{Notations and Definitions}
Let Alice be the sender with state space $\h_A$, and let Bob be the
receiver with output state space $\h_B$. A \textit{quantum channel}
$\E$ is a completely positive map from $\mathcal{B}(\h_A)$ to
$\mathcal{B}(\h_B)$ that can be written into the form
$\E(\rho)=\sum_{k=1}^N E_k\rho E_k^\dagger,$ where $\{E_k:1\leq
k\leq N\}$ is the set of Kraus operators and the completeness
condition $\sum_{k=1}^N E_k^\dagger E_k=I_A$ is satisfied. A
\textit{super-operator} is a completely positive map for which the
completeness condition doesn't need to be satisfied. For simplicity,
sometimes we identify a super-operator $\E$ with Kraus operators
$\{E_k:1\leq k\leq n\}$ by $\E=\{E_k:1\leq k\leq n\}$.

A given quantum channel $\E$ can be used for zero-error
communication as follows: Alice starts with $\ket{0}$, and encodes a
message $k$ into a quantum state
$\rho_k\in\mathcal{B}(\mathcal{H}_{A})$ by a quantum operation
$\E_k$, say $\rho_k=\E_k(\op{0}{0})$. Bob receives $\E(\rho_k)$, and
decodes the message $k$ by suitable quantum operations. Define
$\alpha(\E)$ to be the maximum integer $N$ with which there exist a
set of states $\rho_1,\ldots,\rho_N\in\mathcal{B}(\mathcal{H}_{A})$
such that $\E(\rho_1),\ldots,\E(\rho_N)$ can be perfectly
distinguished by Bob. It follows from the linearity of
super-operators that a set $\{\rho_k: k=1,\ldots,N\}$ achieving
$\alpha(\E)$ can be assumed without loss of generality to be
orthogonal pure states. In \cite{BS07} $\alpha(\E)$ was termed as
{\it the quantum clique number} of $\E$. Intuitively, one use of
$\E$ can be used to transmit $\log_2 \alpha(\E)$ bits of classical
information perfectly. When $\alpha(\E)=1$ it is clear that by a
single use of $\E$ Alice cannot transmit any classical information
to Bob with zero probability of error.

The {\em zero-error classical capacity} of $\E$, $C^{(0)}(\E)$, is
defined as follows:
\begin{equation}\label{c0}
C^{(0)}(\E)=\sup_{k\geq 1}\frac{\log_2 \alpha(\E^{\otimes k})}{k}.
\end{equation}

If we are concerned with the transmission of quantum information
instead of classical information, the notion of zero-error quantum
capacity can be similarly introduced. Let $\alpha^q(\E)$ be the
maximum integer $k$ so that there is a $k$-dimensional subspace
$\h_A'$ of $\h_A$ can be perfectly transmitted through $\E$. That
is, there is a recovery trace-preserving quantum channel
$\mathcal{R}$ from $\mathcal{B}(\h_B)$ to $\mathcal{B}(\h_{A'})$
such that $(\mathcal{R}\circ \E)(\op{\psi}{\psi})=\op{\psi}{\psi}$
for any $\ket{\psi}\in \h_{A'}$. Clearly, the quantity $\log_2
\alpha^q(\E)$ represents the optimal number of intact qubits one can
send by a single use of $\E$. The {\em zero-error quantum capacity}
of $\E$, $Q^{(0)}(\E)$, is defined as follows:
\begin{equation}\label{q0}
Q^{(0)}(\E)=\sup_{k\geq 1}\frac{\log_2 \alpha^q(\E^{\otimes k})}{k}.
\end{equation}
In the following discussion, we mainly focus on the properties of
$\alpha(\E)$ and $C^{(0)}(\E)$.

We will frequently employ the notion of unextendible bases (UB).
Although this notion can be defined on arbitrary multipartite state
space (see Ref. \cite{DFJY07}), for our purpose here it suffices to
focus on matrix spaces. Let $S$ be a set of matrices on
$\mathcal{B}(\h_d)$. $S$ is  said to be a UB if $S^\perp$ contains
no rank-one matrix; otherwise $S$ is said to be extendible. Clearly,
when $S$ is a UB, any nonzero matrix in $S^\perp$ with rank at least
two. In this case we say $S^\perp$ is completely entangled. If $S$
is a UB and can be spanned by rank-one matrix only, we say $S$ an
unextendible product bases (UPB). The properties of UB, in
particular UPB, have been extensively studied in literature. We just
mention two of them here. The first one is that the tensor product
of two UPB is again another UPB \cite{DMS+00}. The second one is
that if the dimension of a matrix subspace $S$ is small enough, say
$\dim(S)< 2d-1$, $S$ is always extendible \cite{PAR04}.
\section{the quantum clique number $\alpha(\cdot)$ is strongly super-multiplicative}
Suppose that $\E$ is classical (a so-called memoryless stationary
channel), that is, $\E=\sum_k \braket{k}{\cdot}{k}\rho_k$ for some
states $\rho_k$ diagonalized under the computational basis
$\{\ket{k}\}$. Then $\alpha(\E)=1$ if and only if for all pairs of
$k$ and $l$, $\rho_k\rho_l\ne0$. Thus $\alpha(\E)=1$ if and only if
$\alpha(\E^{\otimes k})=1$ for any $k$. Therefore, $C^{(0)}(\E)=0$
if and only if $\alpha(\E)=1$. In fact, we can prove that for all
entanglement-breaking channel $\E$ of the form $\E(\rho)=\sum_k
\tr(M_k^\dagger M_k\rho)\rho_k$, where $\{M_k\}$ is a generalized
measurement satisfying $\sum_k M_k^\dagger M_k=I$, it always holds
that $\alpha(\E)=1$ implies that $C^{(0)}(\E)=0$. (See Corollary
\ref{e-b} below for a proof)

We will show that for quantum channels it would be very different.
Let $\E=\sum_{k=1}^n E_k\cdot E_k^\dagger$, where $E_k^\dagger
E_k=I_A$. Let us define
\begin{equation}\label{ke}
\mathcal{K}(\E)={\rm span}\{E_k^\dagger E_l: 1\leq k,l\leq n\}.
\end{equation} $\mathcal{K}(\E)$ plays an
important role in determining the properties of zero-error capacity,
mainly due to the following useful lemma:
\begin{lemma}\label{zero-error}\upshape
Let $\E=\{E_k:1\leq k\leq n\}$ be a quantum channel. Then
$\alpha(\E)>1$ if and only if $\mathcal{K}(\E)$ is extendible, i.e.,
$\mathcal{K}^{\perp}(\E)$ contains a rank-one matrix.
\end{lemma}
{\bf Proof.} Necessity: $\alpha(\E)>1$ implies there are pure states
$\ket{\psi_0}$ and $\ket{\psi_1}$ such that
$\E(\op{\psi_0}{\psi_0})$ and $\E(\op{\psi_1}{\psi_1})$ are
orthogonal. Substituting Kraus sum representation of $\E$ into
$$\tr(\E^\dagger(\op{\psi_0}{\psi_0})\E(\op{\psi_1}{\psi_1}))=0,$$ we
have that $\tr(E_k^\dagger E_l \op{\psi_0}{\psi_1})=0$ for any
$1\leq k,l\leq n$. In other words, $\mathcal{K}(\E)$ is extendible.
Reversing the above arguments we can easily verify the
sufficiency.\hfill$\blacksquare$

Combining the properties of UB mentioned above, we have the
following immediate corollary.
\begin{corollary}\upshape\label{e-b}
Let $\E=\{E_k:1\leq k\leq n\}$ be a quantum channel with input state
space $\mathcal{B}(\h_d)$. Then we have i) If $n<\sqrt{2d-1)}$, then
$\alpha(\E)>1$; ii) If $\mathcal{K}(\E)$ is spanned by a set of
rank-one matrices, then $\alpha(\E)=1$ implies $C^{(0)}(\E)=0$. In
particular, any entanglement-breaking channel satisfies this
property.
\end{corollary}

For any quantum channel $\E$  with  a set of Kraus operators
$\{E_k:1\leq k\leq n\}$ and input state space $\mathcal{B}(\h_d)$,
one can readily verify that $\mathcal{K}(\E)$ satisfies: a)
$\mathcal{K}^\dagger(\E)=\mathcal{K}(\E)$; and b) $I_d\in
\mathcal{K}(\E)$. A somewhat surprising fact is that for a given
matrix subspace $\mathcal{M}\subseteq \mathcal{B}(\h_d)$, these two
properties guarantee the existence of a quantum channel $\E$ such
that $\mathcal{K}(\E)=\mathcal{M}$. Here we define
$\mathcal{M}^\dagger=\{M^\dagger: M\in\mathcal{M}\}$.
\begin{lemma}\label{keylemma}\upshape
Let $\mathcal{M}$ be a matrix subspace of $\mathcal{B}(\h_d)$. Then
there is a quantum channel $\E$ from $\mathcal{B}(\h_d)$ to
$\mathcal{B}(\h_{d'})$ for some integer $d'$ such that
$\mathcal{K}(\E)=\mathcal{M}$ if and only if
$\mathcal{M}^\dagger=\mathcal{M}$ and $I_d\in \mathcal{M}$.
\end{lemma}

{\bf Proof.} Necessity is trivial. We only prove sufficiency.  First
it is easy to see that when $\mathcal{M}^\dagger=\mathcal{M}$, we
can choose a Hermitian basis for $\mathcal{M}$. Actually, for any
matrix $M\in\mathcal{M}$, we know that $M^\dagger\in \mathcal{M}$.
On the other hand, $M$ and $M^\dagger$ can be spanned by two
Hermitian matrices $M+M^\dagger$ and $i(M-M^\dagger)$. So we can
choose a Hermitian basis for $\mathcal{M}$, say $\{M_1,\cdots,
M_n\}$.

Second we show this basis can be made positive definite. Let us
choose a positive real number $s$ and consider $F_k=I_d+s M_k$.
Since $M_k$ is Hermitian, for sufficiently small $s$, all $F_k$ can
be made positive definite. Consider $F_0=I_d-t\sum_{k=1}^n F_k$.
Similarly, choose $t$ sufficiently small we can guarantee that $F_0$
is positive definite. So we have a set of positive definite matrices
$\{F_k: 0\leq k\leq n\}$ such that $\sum_{k=0}^n F_k=I_d$ and ${\rm
span}\{F_k:0\leq k\leq n\}=\mathcal{M}$.

Third, for each operator $F_k$, we will  construct a super-operator
$\E_k$ from $\mathcal{B}(\h_d)$ to $\mathcal{B}(\h_d^{(k)})$, where
$\h_d^{(k)}$ and $\h_d^{(l)}$ are pairwise orthogonal for $0\leq
k\neq l\leq n$. Take the spectral decomposition of $M_k=\sum_{j=1}^d
m_j^{(k)}\op{\psi^{(k)}_j}{\psi^{(k)}_j}$, and let
$\{\ket{j}^{(k)}:1\leq j\leq d\}$ be an orthonormal basis for
$\h_d^{(k)}$. Define a super-operator $\E_k=A_k\cdot A_k^\dagger$,
where
$$A_k=\sum_{j=1}^d
\sqrt{m_j^{(k)}}\op{j^{(k)}}{\psi^{(k)}_j}.$$ It is clear that $A_k$
is from $\mathcal{B}(\h_d)$ to $\mathcal{B}(\h_d^{(k)})$ and
$A_k^\dagger A_k=M_k$.  Now the desired quantum operation $\E$ is
given by the sum of $\E_k$, namely $\E=\sum_{k=0}^n A_k\cdot
A_k^\dagger$. The output space $\h_{(n+1)d}=\oplus_{k=0}^n
\h_{d}^{(k)}$. To prove that $\mathcal{K}(\E)=\mathcal{M}$ one only
needs to notice that $A_k^\dagger A_j=\delta_{kj} M_k$.
\hfill$\blacksquare$

The above lemma greatly simplifies the study of zero-error classical
capacity of noisy quantum channels. It enables us to focus on the
matrix subspaces satisfying two very easily grasped conditions. Some
remarks are as follows:

\begin{enumerate}[(i)]
\item The condition b) ensures that a trace-preserving
super-operator can be found. For our purpose here, we only need
there is a positive definite matrix $M$. Then a super-operator $\E$
with Kraus operators $\{A_k\}$ such that $\sum_{k}A_k^\dagger A_k=M$
can be similarly constructed. Based on $\E$ we can further construct
a trace-preserving quantum operation $\E'$ with Kraus operations
$\{A_k M^{-1/2}\}$. It is easy to check that
$\alpha(\E)=\alpha(\E')$. (Here we assume $\alpha(.)$ is also
defined for any super-operator $\E=\sum_k E_k\cdot E_k^\dagger$ such
that $\sum_k E_k^\dagger E_k$ is positive definite)

\item None of the conditions a) and b) can be further relaxed. This
can be seen from a one-dimensional matrix spanned by a Hermitian
matrix with both negative and positive eigenvalues.

\item In general $\mathcal{M}$ itself may not satisfy conditions a)
and b). However, sometimes we may find two nonsingular matrices $E$
and $F$ so that $\mathcal{M}'=E\mathcal{M}F$ satisfies conditions a)
and b). The extendibility of $\mathcal{M}'$ remains the same as that
of $\mathcal{M}$. That is, for any matrix subspace $\mathcal{M}''$,
$\mathcal{M}\otimes \mathcal{M}''$ is extendible if and only if
$\mathcal{M}'\otimes \mathcal{M}''$ is extendible.

\item After we construct a set of positive semi-definite matrices
$\{M_k\}$ such that $\sum_k M_k=I_d$ and ${\rm
span}\{M_k\}=\mathcal{M}$, we can use a more compact construction of
the corresponding channel $\E$. To do this we introduce an auxiliary
output system $\h_E$ and construct $\E$ from $\mathcal{B}(\h_d)$ to
$\mathcal{B}(\h_d\otimes \h_E)$ as follows:
$$\E(\rho)=\sum_{k=1}^N A_k\rho A_k^\dagger\otimes \op{k}{k},$$
where $A_k=M_k^{1/2}$ is the positive root of $M_k$, and $\ket{k}$
is an orthonormal basis for $\h_E$. Intuitively, $\h_E$ can be
treated as a friendly environment who also outputs its measurement
outcome $k$ after the interaction. One can readily verify that
$\mathcal{K}(\E)={\rm span}\{M_k:1\leq k\leq N\}$. Note that here
the output of $\h_E$ is classical information so that a classical
system is sufficient for our purpose here. This is an example of
quantum communication with classical control.
\end{enumerate}

The following lemma shows that the function of quantum clique number
$\alpha(\cdot)$ is strongly super-multiplicative.
\begin{lemma}\upshape\label{main-lemma}
There are noisy quantum channels $\E$ and $\F$ such that
$\alpha(\E)=\alpha(\F)=1$ and $\alpha(\E\otimes \F)>1$.
\end{lemma}

{\bf Proof.} By Lemmas \ref{zero-error} and \ref{keylemma}, we only
need to construct two unextendible matrix subspaces $S_0$ and $S_1$
both satisfy conditions a) and b), and $S_0\otimes S_1$ are
extendible.

Let $S_0$ be a matrix subspace spanned by the following matrix
bases:
\begin{eqnarray*}\label{S0}
A_1&=&\op{0}{0}+\op{1}{1},\nonumber\\
A_2&=&\op{2}{2}+\op{3}{3},\nonumber\\
A_3&=&\op{2}{0}-\op{0}{2},\nonumber\\
A_4&=&\op{3}{0}+\op{0}{3},\nonumber\\
A_5&=&\op{1}{3}+\op{3}{1},\nonumber\\
A_6&=&\cos\theta\op{0}{1}+\sin\theta\op{2}{3}-\op{1}{2},\nonumber\\
A_7&=&\cos\theta\op{1}{0}+\sin\theta\op{3}{2}-\op{2}{1},\nonumber\\
A_8&=&\sin\theta\op{0}{1}-\cos\theta\op{2}{3}+\sin\theta\op{1}{0}-\cos\theta\op{3}{2},\nonumber
\end{eqnarray*}
where $0<\theta<\pi/2$ is a parameter. Let
$U=\op{0}{0}-\op{1}{1}+\op{2}{2}-\op{3}{3}$, and let $S_1=U
S_0^\perp$, where $S_0^\perp$ is the orthogonal complement via
Hilbert-Schmidt inner product. More explicitly, $S_1$ is spanned by
the following matrix bases:
\begin{eqnarray*}\label{S1}
A_1'&=&\op{0}{0}+\op{1}{1},\nonumber\\
A_2'&=&\op{2}{2}+\op{3}{3},\nonumber\\
A_3'&=&\op{2}{0}+\op{0}{2},\nonumber\\
A_4'&=&\op{3}{0}+\op{0}{3},\nonumber\\
A_5'&=&\op{1}{3}-\op{3}{1},\nonumber\\
A_6'&=&\cos\theta\op{0}{1}+\sin\theta\op{2}{3}-\op{1}{2},\nonumber\\
A_7'&=&\cos\theta\op{1}{0}+\sin\theta\op{3}{2}-\op{2}{1},\nonumber\\
A_8'&=&\sin\theta\op{0}{1}-\cos\theta\op{2}{3}+\sin\theta\op{1}{0}-\cos\theta\op{3}{2},\nonumber
\end{eqnarray*}
We choose $S_1$ as $U S_0^\perp$ instead of $S_0^\perp$ so that
$S_1$ satisfies the Hermitian condition $S_1^\dagger=S_1$ and
contains the identity matrix $I$. This is a key difference from the
previous work \cite{DS08}. By the above lemma, we can define quantum
channels $\E$ and $\F$ such that $\mathcal{K}(\E)=S_0$ and
$\mathcal{K}(\F)=S_1$.

For any $0<\theta<\pi/2$, we will show that $S_0$ and $S_1$ satisfy
the following useful properties:
\begin{enumerate}[(i)]
\item Both $S_0$ and $S_1$ are completely entangled and unextendible.
\item $S_0 \otimes S_1$ are extendible.
\end{enumerate}

Property (ii) holds as $S_0\otimes S_1$ is orthogonal to the
following rank-one element $(I\otimes U)(\op{\Phi_4}{\Phi_4})$,
where $\ket{\Phi_4}=(\ket{00}+\ket{11}+\ket{22}+\ket{33})/2$.

We now prove Property (i). Let $\op{\psi}{\phi}$ be a rank one
matrix orthogonal to $S_0$, where
$\ket{\psi}=\sum_{k=0}^3a_k\ket{k}$ and $\ket{\phi}=\sum_{l=0}^3
b_l^*\ket{l}$. Then we have $\tr(A_k\op{\psi}{\phi})=0$ for $1\leq
k\leq 8$, that is,
\begin{eqnarray*}\label{eqns}
a_0b_0+a_1b_1&=&0,\nonumber\\
a_2b_2+a_3b_3&=&0,\nonumber\\
a_2 b_0- a_0b_2&=&0,\nonumber\\
a_3b_0+a_0b_3&=&0,\nonumber\\
a_1b_3+a_3b_1&=&0,\nonumber\\
\cos\theta a_0b_1+\sin\theta a_2b_3-a_1b_2&=&0,\nonumber\\
\cos\theta a_1b_0+\sin\theta a_3b_2-a_2b_1&=&0,\nonumber\\
\sin\theta a_0b_1-\cos\theta a_2b_3+\sin\theta a_1b_0-\cos\theta
a_3b_2&=&0.
\end{eqnarray*}
Suppose that $a_0b_0\ne 0$. Assume without loss of generality that
$a_0=b_0=1$. Then
\begin{eqnarray*}
a_1b_1=-1,~a_2b_2=-a_3b_3,~a_2=b_2\nonumber \\
a_3=-b_3,~a_1b_3=-a_3b_1.
\end{eqnarray*}
Substituting $a_2=b_2$ and $a_3=-b_3$ into $a_2b_2=-a_3b_3$, we have
$a_2^2=a_3^2$. Similarly substituting $a_1b_1=-1$ and $b_3=-a_3$
into $a_1b_3=-a_3b_1$ we have $a_3(1+a_1^2)=0$. If $a_3=0$ then
$a_2=b_2=b_3=0$. Hence $\cos\theta a_0b_1=\cos\theta
a_0b_1+\sin\theta a_2b_3-a_1b_2=0$, which is a contradiction as both
$a_0$ and $b_1$ are nonzero. Thus $a_1^2=-1$. By $a_1b_1=-1$ we know
that $a_1=b_1=\pm i$. Substituting $b_2=a_2$ and $b_3=-a_3$ into the
last equation we have $\sin\theta b_1+\cos\theta a_2a_3+\sin\theta
b_1-\cos\theta a_3a_2=0$. That is, $2\sin\theta b_1=0$. Again a
contradiction. Therefore $a_0b_0=0$. Note that if $a_kb_l=0$ and for
a nonzero constant $\lambda$, $\lambda a_kb_{l'}=a_{k'}b_{l}$, then
$a_{k}b_{l'}=a_{k'}b_{l}=0$. Applying this inference rule many
times, one concludes that all $a_kb_l=0$, $0\le k, l\le 3$ in both
$a_0=0$ and $b_0=0$ cases. Thus $\op{\psi}{\phi}=0$, and $S_0$ is
unextendible. By the same technique, we can prove that $S_1$ is also
unextendible.

Applying Lemma \ref{zero-error}, we know that
$\alpha(\E)=\alpha(\F)=1$. On the other hand, by property (ii) we
know that $\alpha(\E\otimes \F)\ge 2$. Actually, Alice can use
$\ket{\Psi_0}=\ket{\Phi_4}$ and $\ket{\Psi_1}=(I\otimes
U)\ket{\Phi_4}$ to encode ``0" and ``1", respectively, and Bob can
recover this bit by distinguish between $(\E\otimes
\F)(\op{\Psi_0}{\Psi_0})$ and $(\E\otimes \F)(\op{\Psi_1}{\Psi_1})$,
which are orthogonal by our construction. \hfill$\blacksquare$

In the above construction, $\E$ and $\F$ are not identical. However,
using the direct sum construction \cite{FW07}, we can find a quantum
channel enjoying similar property. Now we are ready to present our
main result:
\begin{theorem}\upshape \label{th1}
There is a classical of quantum channels $\G$ such that
$\alpha(\G)=1$  and $\alpha(\G^{\otimes 2})>1$. Hence
$C^{(0)}(\G)\geq 0.5$.
\end{theorem}
{\bf Proof.} The idea is to take $\G$ as the direct sum of $\E$ and
$\F$, say $\G=\E\oplus\F$. More explicitly,
$$\G(\rho)=\E(P_0\rho P_0)+\F(P_1\rho P_1),$$
where $P_0$ and $P_1$ are the projections on the input state spaces
of $\E$ and $\F$, respectively, and $P_0+P_1$ is the projection of
the whole input state space for $\G$. The function of $\G$ can be
understood as follows: for any input state $\rho$, we first perform
a projective measurement $\{P_0, P_1\}$. If the outcome is $0$, then
we apply $\E$ to the resulting state; otherwise we apply $\F$. It is
clear that $\alpha(\G)=\max\{\alpha(\E),\alpha(\F)\}$. Furthermore,
we have
\begin{equation*}
\alpha(\G^{\otimes 2})=\max\{\alpha(\E^{\otimes 2}),\alpha(\E\otimes
\F),\alpha(\F\otimes \E),\alpha(\F^{\otimes 2})\}.
\end{equation*}
For channels $\E$ and $\F$ constructed above, we have $\alpha(\G)=1$
and $\alpha(\G^{\otimes 2})\geq
\alpha(\E\otimes\F)>1$.\hfill$\blacksquare$

Based on our previous work about UB \cite{DCX09}, we know that any
channel $\G$ with the property in Theorem \ref{th1} should be at
least with a $4$-dimensional input state space. It remains unknown
how to construct quantum channel with similar property and with
smaller input and output dimensions.

\section{A class of special quantum channels}
The construction in Lemma \ref{keylemma} suggests us to consider a
special class of quantum channels, which can be treated as a
generalization of retro-correctible channels introduced in
\cite{BDSS06}. Consider a quantum channel $\E$ from
$\mathcal{B}(\h_c\otimes \h_d)$ to $\mathcal{B}(\h_{c'}\otimes
\h_{d'})$ as follows:
$$\E=\sum_{k=1}^N \E_k\otimes \F_k,$$
where both $\E_k$ and $\F_k$ are super-operators for each $1\leq
k\leq N$. Usually we choose $\{\F_k\}$ to be a set of quantum
channels and $\{\E_k\}$ is a set of super-operators such that
$\sum_{k=1}^N \E_k$ is trace-preserving.
\begin{figure}[ht]
  \centering
  \includegraphics[scale=0.5]{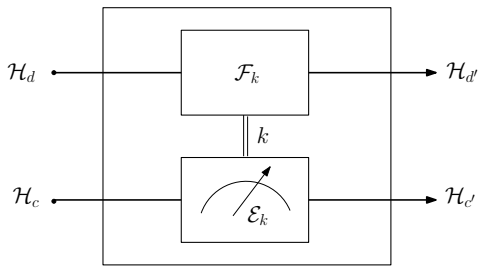}
  \caption{
  Internal realization of a controlled communication channel $\E$:
1) Perform a measurement $\{\E_k\}$ to the control input system
$\h_c$; 2) If the measurement outcome is $k$, apply $\F_k$ to the
data input $\h_d$; 3) Output both the control and the data inputs to
$\h_{c'}$ and $\h_{d'}$, respectively. Here the input dimensions $c$
and $d$ are not required to be the same as the output dimensions
$c'$ and $d'$, respectively.}
  \label{pic2}
\end{figure}

Imposing special constraints on $\E_k$ and $\F_k$, we can construct
some quantum channels with desirable properties. In particular, if
the receiver Bob can distinguish between $\{\E_k: 1\leq k\leq N\}$,
he will be able to determine the quantum operation performed on the
data input exactly. Thus the net effect of the channel $\E$ will
reduce to some of $\F_k$. In the case that $\F_k$ has a large amount
of classical or quantum capacity, the above channel will also have
large capacity. Symmetrically, if Bob can distinguish between $\F_k$
then he will be able to know the measurement operator performed by
the environment, and then is able to correct the errors.

For example, if we choose $d'=c'=c=n$, and $d=1$, and choose $\F_k$
to be the unitary (isometry) $\op{k}{0}$ from $\h_d$ to $\h_{d'}$,
and let control operators $\E_k$ be a set of generalized measurement
$\{A_k:1\leq k\leq n\}$ from $\h_{c}$ to $\h_{c'}$, then we have the
following channel:
$$\E(\rho\otimes \op{0}{0})=\sum_{k=1}^{n}(A_k\otimes \op{k}{0})(\rho \otimes \op{0}{0})(A_k^\dagger\otimes \op{0}{k}).$$
If we ignore the one-dimensional data input, the above channel can
be simplified as follows:
$$\E(\rho)=\sum_{k=1}^{n}A_k \rho A_k^\dagger\otimes \op{k}{k}.$$
This is precisely the channel we introduced in the previous section,
where a similar interpretation has been presented.

Another special case is that $\{\F_k\}$ or $\{\E_k\}$ are not
distinguishable in general, but would be distinguishable if an
entangled state $\ket{\Phi}$ is provided. That is, the set of
$\{(I\otimes \E_k)(\op{\Phi}{\Phi})\}$ is distinguishable.
Intuitively, $\{\E_k\}$ cannot be distinguishable means that the
channel is very noisy. So the capacity without any assistance would
be generally small. However, supplying additional resources such as
shared entanglement will greatly improve the capacity. The class of
retro-correctible channels introduced by Bennett et al \cite{BDSS06}
is a typical example.

\section{Super-Additivity of Zero-error classical and quantum capacities}
It was known that the zero-error classical capacity of classical
channels are super-additive in the following sense
\cite{LOV79,HAE79}: there are classical channels $\mathcal{N}_0$ and
$\mathcal{N}_1$ such that $C^{(0)}(\mathcal{N}_0\otimes
\mathcal{N}_1)>C^{(0)}(\mathcal{N}_0)+C^{(0)}(\mathcal{N}_1)$. This
is very different from the ordinary capacity, which is always
additive. However, any classical channels $\mathcal{N}_0$ and
$\mathcal{N}_1$ satisfying the super-additivity must have the
ability to transfer classical information perfectly, that is
$C^{(0)}(\mathcal{N}_0)>0$ and $C^{(0)}(\mathcal{N}_1)>0$.  It
remains unknown whether the above super-additivity still holds if
one or two quantum channels are with vanishing zero-error capacity.
Here we will show that both $C^{(0)}$ and $Q^{(0)}$ satisfy a
stronger type of super-additivity. Let's consider $C^{(0)}$ first.
\begin{theorem}\upshape\label{th2}
There is an entanglement-breaking channel $\mathcal{E}$ on
$\mathcal{B}(\h_{2d})$ such that $C^{(0)}(\E)=0$ and
$C^{(0)}(\mathcal{I}_2\otimes \E)=\log_2 d\gg
C^{(0)}(\mathcal{I}_2)+C^{(0)}(\E)=1$, where $\mathcal{I}_2$ is one
qubit noiseless quantum channel.
\end{theorem}

Proof. Consider the quantum channel
$$\E=\E_0\otimes\F_0+\E_1\otimes \F_1,$$
where $\E_0=\{\op{0}{0},\op{1}{+}\}$,
$\E_1=\{\op{0}{1},\op{1}{-}\}$, $\F_0=\{\op{k}{k}: 1\leq k\leq d\}$
and $\F_1=\{\op{\bar{k}}{k}:1\leq k\leq d\}$. In particular,
$\{\ket{k}\}$ and $\{\ket{\bar{k}}\}$ are two orthonormal bases such
that $\ip{k}{\bar{l}}\neq 0$. By choice we have that
$\F_0(\rho)\perp \F_1(\sigma)=0$ if and only if $\rho=0$ or
$\sigma=0$. It is also clear that $\alpha(\F_0)=\alpha(\F_1)=d$, and
the set of input states can be chosen as $\{\ket{k}:1\leq k\leq
d\}$. These facts will be useful later.

First we show that $C^{(0)}(\E)=0$. Clearly $\E$ is an entanglement
breaking channel as it has a set of rank-one Kraus operators. Thus
it suffices to show that $\alpha(\E)=1$. Take an input state
$\ket{\psi}=\ket{0}\ket{\psi_0}+\ket{1}\ket{\psi_1}$ and calculate
\begin{eqnarray*}
\E(\psi)&=&\op{0}{0}\otimes(\F_0(\psi_0)+\F_1(\psi_1))\\
        &+&\op{1}{1}\otimes(\F_0(\psi_0+\psi_1)+\F_1(\psi_0-\psi_1)),
\end{eqnarray*}
where for simplicity we assume $\psi_0=\op{\psi_0}{\psi_0}$ and
$\psi_0+\psi_1=\op{\psi_0+\psi_1}{\psi_0+\psi_1}$, etc.  Similarly,
for another input state
$\ket{\phi}=\ket{0}\ket{\phi_0}+\ket{1}\ket{\phi_1}$, we have
\begin{eqnarray*}
\E(\phi)&=&\op{0}{0}\otimes (\F_0(\phi_0)+\F_1(\phi_1))\\
        &+&\op{1}{1}\otimes(\F_0(\phi_0+\phi_1)+ \F_1(\phi_0-\phi_1)).
\end{eqnarray*}

If $\E(\psi)$ and $\E(\phi)$ are orthogonal, we should have
\begin{eqnarray}
\F_0(\psi_0)\F_1(\phi_1)=0,\\
\F_1(\psi_1)\F_0(\phi_0)=0,\\
\F_0(\psi_0+\psi_1)\F_1(\phi_0-\phi_1)=0,\\
\F_1(\psi_0-\psi_1)\F_0(\phi_0+\phi_1)=0.
\end{eqnarray}
From the first equation we know that $\psi_0=0$ or $\phi_1=0$.
Without loss of generality, assume that $\psi_0=0$. It follows from
the second equation $\phi_0=0$ as $\psi_1\neq 0$. However this would
imply that both $\psi_0+\psi_1=\psi_1$ and $\phi_0-\phi_1=-\phi_1$
are nonzero, thus the third equation cannot hold. With that we
complete the proof of $\alpha(\E)=1$.

The next step is to show that if a noiseless qubit channel
$\mathcal{I}_2$ is supplied between Alice and Bob, Alice can send
$d$ messages perfectly to Bob using $\mathcal{I}_2\otimes \E$. Let
$\ket{\Phi_2}=(\ket{00}+\ket{11})/\sqrt{2}$. The key here is that
$\E_0$ and $\E_1$ are distinguishable by $\ket{\Phi_2}$ in the sense
that
$$\rho_0=(I_2\otimes
\E_0)(\Phi_2)=(\op{00}{00}+\op{+1}{+1})/2$$ and
$$\rho_1=(I_2\otimes \E_1)(\Phi_2)=(\op{10}{10}+\op{-1}{-1})/2$$ are
orthogonal. If Alice encodes message $k$ into $\ket{\Phi_2}\otimes
\ket{k}$ and transmits it to Bob via $\mathcal{I}_2\otimes \E$, the
received states by Bob are
$$\{(\rho_0\otimes \op{k}{k}+\rho_1\otimes \op{\bar{k}}{\bar{k}})/2:1\leq k\leq d\},$$
which are mutually orthogonal. That completes the proof of
$\alpha(\mathcal{I}_2\otimes \E)\geq d$. \hfill$\blacksquare$

It is easy to see that the role of the noiseless qubit channel
$\mathcal{I}_2$ can be replaced with a pre-shared $2\otimes 2$
maximally entangled state $\ket{\Phi_2}$ between Alice and Bob. To
encode the message $k$, Alice simply sends $\ket{k}$ together with
her half of entangled state to Bob. The received states by Bob are
the same as above.

With a more careful analysis we can easily see that $\rho_0$ and
$\rho_1$ are locally distinguishable in the following way: Bob
performs a projective measurement according to
$\{\ket{0},\ket{1}\}$, and then sends outcome $b$ to Alice. If $b=0$
then Alice measures her particle using the same basis, otherwise
using diagonal basis $\{\ket{+},\ket{-}\}$. The outcomes $00,+1$
correspond to $\rho_0$, while $10, -1$ correspond to $\rho_1$. So a
more economic way to achieve the perfect transmission is that: Alice
locally prepares a Bell state $\ket{\Phi_2}$ and then send $\ket{k}$
and one half of $\ket{\Phi_2}$ to Bob. Bob feedbacks his measurement
outcome on the control qubit to Alice. Based on Bob's information,
Alice performs the measurement on the left half of $\ket{\Phi_2}$
and forwards the measurement outcome to Bob. Bob will then know
which of $\F_0$ and $\F_1$ is performed on the data input and can
perfectly decode the message $k$. Note here we use two-way classical
communication which is usually not allowable. However, from the
above analysis we can see these communications are independent from
the message $k$ we send in our main protocol.  To summarize, we have
the following

\begin{corollary}\upshape
For the quantum channel $\E$ constructed in above theorem, we have
1) $C^{(0)}_{1~ebit}(\E)\geq \log_2 d \gg C^{(0)}(\E)=0$, where the
subscript means one ebit available; 2) $C^{(0)}_2(\E)\geq \log_2 d
\gg C^{(0)}(\E)=0$, where the subscript $2$ denotes the two-way
classical communication independent of the message sending through
the main protocol.
\end{corollary}

So far we haven't touched the zero-error quantum capacity yet. Using
a similar construction, we can prove the strong super-additivity of
$Q^{(0)}$. A somewhat surprising fact is that even for quantum
channel with vanishing zero-error classical capacity, the
super-activation effect remains possible. Actually we have the
following
\begin{theorem}\upshape\label{th3}
There is quantum channel $\mathcal{E}$ with input state space
$\mathcal{B}(\h_{2d})$ such that $C^{(0)}(\E)=0$ and
$Q^{(0)}(\mathcal{I}_2\otimes \E)=\log_2 d\gg
Q^{(0)}(\mathcal{I}_2)+ Q^{(0)}(\E)=1$, where $\mathcal{I}_2$ is the
noiseless qubit channel.
\end{theorem}

\textbf{Outline of Proof}. Consider the following quantum channel:
\begin{equation}\label{super}
\E=\frac{1}{\sqrt{N}}\sum_{k=1}^N (\E_{k0}\otimes I_d+\E_{k1}\otimes
U_k)
\end{equation}
where $\E_{k0}=\{\op{k}{\psi_{k0}^*}\}$,
$\E_{k1}=\{\op{k}{\psi_{k1}^*}\}$,
$\{\ket{\psi_{k0}},\ket{\psi_{k1}}\}$ is an orthogonal basis for
$\h_2$, ``$*$'' is the complex conjugate according to
$\{\ket{0},\ket{1}\}$, and $\{U_k\}$ is a set of unitary operations
on $\h_d$. The function of $\E$ can be understood as follows: First,
randomly choose an integer $k\in \{1,\cdots, N\}$, and perform a
projective measurement $\{\ket{\psi_{k0}^*},\ket{\psi_{k1}^*}\}$ on
the control input qubit. If the outcome is $0$ no action to the data
input; otherwise perform $U_k$. Second, output the classical
information $k$ but keep the measurement outcome hidden. This is
exactly one special instance of retro-correctible channel
\cite{BDSS06}.

It is easy to see that $\mathcal{K}(\E)$ is given by
\begin{equation*}
\{{\psi_{k0}}\otimes I_d, {\psi_{k1}}\otimes I_d,
\op{\psi_{k0}}{\psi_{k1}}\otimes U_k,
\op{\psi_{k1}}{\psi_{k0}}\otimes U_k^\dagger\},
\end{equation*}
where $1\leq k\leq N$ and recall that
$\psi_{k0}=\op{\psi_{k0}}{\psi_{k0}}$. We can see that
$$\mathcal{K}^\perp(\E)\subseteq \{I_2\otimes D:\tr(D)=0,
D\in\mathcal{B}(\h_d)\}.$$ We will show that by choosing $U_k$,
$\ket{\psi_{k0}}$, $N$ appropriately, the above inequality holds
with equality. To achieve this, we only need to choose
$\ket{\psi_{0k}}$ and $U_k$ so that ${\psi_{k0}}\otimes I_d$ spans
$\mathcal{B}(\h_2)\otimes I_d$, and
$\op{\psi_{k0}}{\psi_{k1}}\otimes U_k$ spans ${\rm span}\{C\otimes
\mathcal{B}(\h_d): \tr(C)=0, C\in \mathcal{B}(\h_2)\}$. This can be
done easily as $${\rm span}\{\op{\psi}{\phi}:
\ip{\psi}{\phi}=0\}=\{D: \tr(D)=0\}$$ and $${\rm span}\{U: U^\dagger
U=I_d\}=\mathcal{B}(\h_d).$$

Now it is clear that $\mathcal{K}(\E)$ contains the following set of
rank-one matrices:
$$\{\op{0}{1},\op{1}{0}, \op{+}{-}\}\otimes \{\op{k}{l}:1\leq k,l\leq d\},$$
which is clearly a UPB as its orthogonal complement is a completely
entangled subspace $I_2\otimes \mathcal{B}(\h_d)$. Thus
$\mathcal{K}(\E^{\otimes n})$ is unextendible for any $n\geq 1$,
which follows $C^{(0)}(\E)=0$.

The argument that the above channel is able to communicate quantum
information perfectly is similar to  the analysis for the
retro-correctible channels. The key idea is that with the assistance
of shared entanglement, the hidden measurement outcome can be
revealed. Suppose that $\ket{\Phi_2}=(\ket{00}+\ket{11})/\sqrt{2}$
is supplied to Alice and Bob. Then by inputting an arbitrary state
$\ket{\psi}$ into the data slot, and Alice's half of the
$\ket{\Phi_2}$ into control slot, Bob receives the following output
state
$$(\mathcal{I}_2\otimes\E)({\Phi_2}\otimes
{\psi})=1/N\sum_{k=1}^{2N}({k\psi_{k0}}\otimes
{\psi}+{k\psi_{k1}}\otimes U_k{\psi} U_k^\dagger).$$  Note that
$\cup_{k=1}^N\{\ket{k\psi_{k0}}, \ket{k\psi_{k1}}\}$ is an
orthonormal basis for $\h_N\otimes\h_2$. Thus Bob can perfectly
distinguish them by a projective measurement. If the outcome is
$k0$, then the data output is ${\psi}$. If the outcome is $k1$, then
the data output is $U_k\psi U_k^\dagger$. Applying $U_k^\dagger$ to
the data output, we can recover ${\psi}$. That means $\E$ together
with one ebit can be used to perfectly transfer a $d$-dimensional
quantum system, or in other words, with entanglement-assisted
zero-error quantum capacity at least $\log_2 d$ qubits.

It is not difficult to see that the role of shared entanglement can
be replaced by a noiseless qubit channel. That immediately implies
$Q^{(0)}(\mathcal{I}_2\otimes \E)\geq \log_2
d>Q^{(0)}(\mathcal{I}_2)+Q^{(0)}(\E)=1$.

Notice further that $\cup_{k=1}^N\{\ket{k\psi_{k0}},
\ket{k\psi_{k1}}\}$ is an orthonormal basis for $\h_N\otimes\h_2$
that is LOCC distinguishable. So the assistance of two-way classical
communications that are independent from the quantum information
sending through the main protocol can be used to transmit $\log_2 d$
noiseless qubits. The analysis is similar to the previous theorem
and we omit the details here.\hfill$\blacksquare$

\begin{corollary}\upshape
There is quantum channel $\E$ with input state space
$\mathcal{B}(\h_{2d})$ such that $C^{(0)}(\E)=0$ and  1)
$C^{(0)}(\mathcal{I}_2\otimes \E)\geq \log_2 d\gg
C^{(0)}(\mathcal{I}_2)+C^{(0)}(\E)=1$; 2)
$Q^{(0)}(\mathcal{I}_2\otimes \E)\geq \log_2 d\gg
Q^{(0)}(\mathcal{I}_2)+Q^{(0)}(\E)=1$; 3) $C^{(0)}_2(\E)\geq
Q^{(0)}_2(\E)\geq \log_2 d$, where the subscript $2$ denotes two-way
classical communications that are independent of the message sending
through the main protocol; 4) $C^{(0)}_{1 ~ebit}(\E)\geq Q^{(0)}_{1
~ebit}(\E)\geq \log_2 d$, where the subscript means one ebit
available.
\end{corollary}

The above corollary indicates the behaviors of zero-error capacity
of quantum channels is very weird: There are quantum channels which
have a large amount quantum capacity but with vanishing zero-error
classical capacity. However, the channel can be unlocked for
zero-error quantum communication if a small amount of additional
resources such as two-way classical communication independent of the
messages sending through the main protocol, shared entanglement, or
a noiseless quantum channel is available.

\section{Classical feedback enables perfect transmission of information}
In this section we will study the role of classical feedback. A well
known result in classical information theory is that a noiseless
classical feedback channel from Bob to Alice cannot increase the
capacity of a classical channel. For quantum channels, it remains an
open problem whether a classical feedback can strictly increase the
capacity \cite{BDSS06}. However, in the special case that a quantum
channel with zero classical capacity  it should be a constant
channel, i.e., it sends any input quantum state to a fixed state.
Clearly, the classical feedback cannot increase the capacity under
this special assumption. The situation is very different for quantum
capacity, which can be increased by classical feedback even the
unassisted quantum capacity is zero. A typical example is the
quantum erasure channel with erasure probability more than $0.5$,
which has vanishing quantum capacity but nonzero classical feedback
assisted quantum capacity \cite{BDS97}.

It was pointed out in \cite{SHA56} that for certain classical noisy
channels a noiseless classical feedback channel from Bob to Alice
may strictly increase the zero-error classical capacity. All these
channels should satisfy $C^{(0)}>0$. In other words, without any
assistance they can be used to communicate classical information
perfectly. Thus a question of interest is to ask whether this
assumption can be removed. We provide an affirmative answer to this
question as follows:
\begin{theorem}\upshape\label{cfb}
There is quantum channel $\E$ such that $C^{(0)}(\E)=0$ but
$C_{cfb}^{(0)}(\E)>0$ and $Q_{cfb}^{(0)}(\E)>0$, where the subscript
``cfb" represents classical feedback from Bob to Alice.
\end{theorem}

\textbf{Proof.} The quantum channel constructed in Eq. (\ref{super})
is exactly one such channel when $d=2$. To see this, one only needs
to show that one use of $\E$ together with back communication can
generate a shared entangled state $\ket{\Phi_d}$.  The protocol is
as follows. First Alice prepares $\ket{\Phi_2}\ket{\Phi_d}$ and send
half of them to Bob. Second Bob measures the control output and
feedbacks the outcome $k$. For the moment he has already known the
shared entangled state between him and Alice should be one of
$\ket{\Phi_d}$ or $(I\otimes U_k)\ket{\Phi_d}$. After receiving $k$,
Alice performs a measurement according to $\{\ket{\psi_{k0}},
\ket{\psi_{k1}}\}$. If $k0$ is obtained, the final shared entangled
state is $\ket{\Phi_d}$; if $k1$ is obtained, the final shared
entangled state is $(I\otimes U_k)\ket{\Phi_d}$, and she only needs
to perform $U_k^*$ to the left half of $\ket{\Phi_d}$, thus the
final resulting state is again $\ket{\Phi_d}$. If $d=2$, we already
know that $\E$ together with this entangled state can be used to
send one noiseless qubit. In total, two uses of $\E$ and classical
feedback enable one noiseless qubit transmission. Therefore
$$C_{cfb}^{(0)}(\E)\geq Q_{cfb}^{(0)}(\E)\geq 0.5.$$

If in Eq. (\ref{super}) we use a $d$-dimensional control input
instead of a $2$-dimensional one, we will know that $C^{(0)}(\E)=0$
but
$$C_{cfb}^{(0)}(\E)\geq
Q_{cfb}^{(0)}(\E)\geq 0.5 \log_2 d.$$

There is, however, a quantum channel with only a two-dimensional
input state space enjoying the same property. Due to its simplicity,
let us give a detailed analysis here. Consider the following
$2\times 2$ matrix subspace
$$\mathcal{M}={\rm span}\{I,X,Y\}=\{Z\}^\perp,$$ where $X, Y, Z$ are
$2\times 2$ Pauli matrices. By Lemma \ref{keylemma}, we know there
is a quantum channel $\E$ such that $\mathcal{K}(\E)=\mathcal{M}$.
We can construct one such channel by the arguments in the proof of
Lemma \ref{keylemma}. Here we will carefully construct one
satisfying special requirement. First choose a positive definite
bases for $\mathcal{M}$ such that $\mathcal{M}={\rm span}\{A_1, A_2,
A_3,A_4\}$ and
\begin{eqnarray*}
A_1&=&(\frac{2}{3}\op{+}{+}+\frac{1}{3}\op{-}{-})/2,\\
A_2&=&(\frac{1}{3}\op{+}{+}+\frac{2}{3}\op{-}{-})/2,\\
A_3&=&(\frac{2}{3}\op{i_+}{i_+}+\frac{1}{3}\op{i_-}{i_-})/2,\\
A_3&=&(\frac{1}{3}\op{i_+}{i_+}+\frac{2}{3}\op{i_-}{i_-})/2,
\end{eqnarray*}
where $\ket{i_{\pm}}=(\ket{0}\pm i\ket{1})/\sqrt{2}$. We have chosen
$A_k$ such that $\sum_{k=1}^4 A_k=I$. Take $E_k=A_k^{1/2}$ and
construct the following quantum channel from $\mathcal{B}(\h_2)$ to
$\mathcal{B}(\h_2\otimes \h_4)$:
$$\E(\rho)=\sum_{k=1}^4 E_k\rho E_k^\dagger \otimes \op{k}{k},$$
where $\{\ket{k}:1\leq k\leq 4\}$ is an orthonormal basis for
auxiliary system.

Noticing that $\mathcal{K}(\E)={\rm
span}\{\op{+}{+},\op{-}{-},\op{i_+}{i_+}\}$ is a UPB, we have that
$\alpha(\E^{\otimes n})=1$ for any $n\geq 1$. Thus $C^{(0)}(\E)=0$.
On the other hand, $\mathcal{K}^\perp(\E)=Z$. So if a maximally
entangled state $\ket{\Phi_2}$ is shared between Alice and Bob,
Alice can send one bit to Bob without any error. To do this, Alice
first encodes ``0" by applying $I$ and ``1" by applying $Z$ to her
half of the shared entangled state, respectively, and sends her half
of $\ket{\Phi_2}$ to Bob. The received states by Bob are
$$\rho_0=\sum_{k=1}^4 (I\otimes E_k)\op{\Phi_2}{\Phi_2}(I\otimes E_k)^\dagger\otimes \op{k}{k},$$
and
$$\rho_1=\sum_{k=1}^4(I\otimes E_kZ)\op{\Phi_2}{\Phi_2}(I\otimes E_k Z)^\dagger \otimes \op{k}{k},$$
respectively. By our assumption on $E_k$, $\rho_0$ and $\rho_1$ are
orthogonal. Thus Bob can decode the bit perfectly.

Now the whole problem is reduced to generate a maximally entangled
state between Alice and Bob using $\E$ and classical feedback only.
Fortunately, this can be done as follows:

Step 1. Alice locally prepares $\ket{\Phi_2}$ and sends one half of
$\ket{\Phi_2}$ to Bob through $\E$.

Step 2. Bob measures the auxiliary system according to
$\{\ket{k}:1\leq k\leq 4\}$. If the outcome is $k$, Bob will know
that an entangled state $\ket{\Psi_k}=\sqrt{2}(I\otimes
E_k)\ket{\Phi_2}$ with Schmidt coefficient vector $(2/3, 1/3)$ is
generated between him and Alice.

Step 3. Repeat steps $1$ and $2$ once more, Alice and Bob will share
a state $\ket{\Psi_k}\otimes \ket{\Psi_l}$, with Schmidt coefficient
vector $(4/9,2/9,2/9,1/9)$. (However only Bob knows the exact form
of $\ket{\Psi_k}\ket{\Psi_l}$ as Alice doesn't know the measurement
outcomes $k$ and $l$).

Step 4. Bob feedbacks the measurement outcomes $k$ and $l$ to Alice.
So Alice also knows the exact form of the shared entangled state
between them.

Step 5. Bob and Alice transform the shared entangled state into a
Bell state with standard form $\ket{\Phi_2}$. By Nielsen's theorem
\cite{NIE99}, this can be achieved with certainty as
$(4/9,2/9,2/9,1/9)\prec (1/2,1/2)$. Furthermore, the transformation
can be done using local measurements and classical communications
from Bob to Alice only.

Combining the above discussions, we know that $3$ uses together with
back communication can transmit one bit perfectly from Alice to Bob.
Thus $C_{cfb}^{(0)}\geq 1/3>0$. Moreover, we can send a qubit by
sending two bits and consuming one ebit. Easily see that $8$ uses of
$\E$ can transmit one noiseless qubit. Hence $Q_{cfb}^{(0)}\geq
1/8$.\hfill$\blacksquare$

It seems that the retro-correctible channel $\mathcal{R}_{2,2}$
introduced in \cite{BDSS06} might enjoy the same property as above.
However, we don't know how to determine the value of
$C^{(0)}(\mathcal{R}_{2,2})$ and consequently, it remains unknown
whether $C^{(0)}(\mathcal{R}_{2,2})$ is  vanishing or not.

\section{Conclusions and Discussions}
In sum, we have demonstrated that for a class of quantum channels, a
single use of the channel cannot be used to transmit classical
information with zero probability of error, while multiple uses can.
This super-activation property is enabled by quantum entanglement
between different uses, thus cannot be achieved by classical
channels. We also have shown that additional resources such as
classical communications independent of sending messages, shared
entanglement, and noiseless quantum communication would be greatly
improve the zero-error capacity for certain channels. In particular,
both the zero-error classical capacity and zero-error quantum
capacity are strongly super-additive even one of the channels is
with vanishing zero-error classical capacity. Finally we construct a
special class of quantum channels to show that the classical
feedback enables perfect transmission of both classical and quantum
information even when the quantum channel has vanishing zero-error
classical capacity.  These results suggest that a new quantum
zero-error information theory would be highly desirable.

Many interesting problems remain open, and here we mention two of
them. The first one is to show whether the following strongest
super-additivity is possible: Find quantum channels $\E$ and $\F$
such that $C^{(0)}(\E)=C^{(0)}(\F)=0$ and $C^{(0)}(\E\otimes \F)>0$.
According to Lemma \ref{keylemma}, this is equivalent to find two
matrix subspaces $S_0$ and $S_1$  such that 1) $I\in S_k$ and
$S_k^\dagger =S_k$; 2) $S_k^{\otimes n}$ are unextendible for any
$n\geq 1$, $k=0,1$; and 3) $S_0\otimes S_1$ are unextendible. The
quantum channels presented in Lemma \ref{main-lemma} may be eligible
candidates. However, we are not able to answer this question at
present as we don't have a feasible way to check whether
$S_k^{\otimes n}$ is extendible for $n>1$. The second one is to
study corresponding problems about the zero-error quantum capacity
$Q^{(0)}$. In this case we don't even know whether $\alpha^q(\cdot)$
is super-multiplicative. A result similar to Lemma \ref{main-lemma}
would be highly desirable. All these problems can be successfully
solved for another notion of unambiguous capacity, which is a
generalization of zero-error capacity by requiring the decoding
process to be unambiguous \cite{DCX09}.

{\bf Note Added:} After the completion of this work, the author
happened to know that Cubitt, Chen, and Harrow also independently
obtained some super-activation results about the zero-error
classical capacity which partially overlap with ours \cite{CCH09}.
More precisely, they employed the Choi-Jamio{\l}kowski isomorphism
between quantum channels and a class of bipartite mixed states to
establish a theorem similar to Lemma \ref{keylemma} here. Then two
quantum channels $\E$ and $\F$ with four-dimensional input state
spaces such that $\alpha(\E)=\alpha(\F)=1$ and $\alpha(\E\otimes
\F)>1$ were explicitly constructed. They further applied some
powerful techniques from Algebraic Geometry to show that a pair of
quantum channels satisfying the strongest super-additivity does
exist, and thus solved the open problem mentioned above. (One of
these techniques is a result about strongly unextendible bases that
was previously proven and used in \cite{DCX09} to demonstrate a
similar super-activation effect for unambiguous capacity of quantum
channels) Clearly, their remarkable result established the strongest
type of super-additivity, which they termed as the super-activation
of the asymptotic zero-error classical capacity of quantum channels.
Interestingly, it is not difficult to show that all channels we
constructed in Theorems \ref{th2}-\ref{cfb} cannot be activated by
any quantum channel $\F$ with $\alpha(\F)=1$. Thus it is still a
surprising fact that these channels do satisfy certain type of
super-activation effects which are definitely impossible for any
classical channel.

\section*{Acknowledgements} Part of this work was completed when
the author was visiting the University of Michigan, Ann Arbor. Many
thanks were given to Yaoyun Shi for his hospitality and for many
inspiring discussions. The author was also grateful to John Smolin
and Graeme Smith for discussions about zero-error capacity during
their short visit to Tsinghua University. In particular, the quantum
channel presented in Eq. (\ref{super}) was very much motivated by
Smolin's suggestions. Delightful discussions with Prof. Mingsheng
Ying, Yuan Feng, Jianxin Chen, and Yu Xin were sincerely
acknowledged. Special thanks were given to T. Cubitt, J. X. Chen,
and A. Harrow for sending me a copy of their manuscript prior to
publication and for useful discussions.  This work was partially
supported by the National Natural Science Foundation of China (Grant
Nos. 60702080, 60736011) and the Hi-Tech Research and Development
Program of China (863 project) (Grant No. 2006AA01Z102). This work
was also partially supported by the National Science Foundation of
the United States under Awards~0347078 and 0622033.


\begin{thebibliography}{99}
\bibitem{SHA56} C. E. Shannon, IRE Trans. Inf. Theory \textbf{2}, 8 (1956).

\bibitem{KO98} J. K$\ddot{\rm o}$rner and A. Orlitsky, IEEE Trans.
Inf. Theory \textbf{44}, 2207 (1998).

\bibitem{LOV79} L. Lov$\acute{\rm a}$sz, IEEE Trans. Inf. Theory \textbf{29}, 1
(1979).

\bibitem{HAE79} W. Haemers, IEEE Trans. Inf. Theory \textbf{25}, 231 (1979).

\bibitem{ALO98} N. Alon, Combinatorica \textbf{18}, 301 (1998); N. Alon and E.
Lubetzky, IEEE Trans. Inf. Theory \textbf{52}, 2172 (2006).

\bibitem{CRST06}  M. Chudnovsky, N. Robertson, P. D. Seymour, and R. Thomas,
Ann. Math. \textbf{164}, 51 (2006).

\bibitem{MA05} R. A. C. Medeiros and F. M. de Assis, Int. J. Quant. Inf.
\textbf{3}, 135 (2005); R. A. C. Medeiros, R. Alleaume, G. Cohen,
and F. M. de Assis, quant-ph/0611042.

\bibitem{BS07} S. Beigi and P. W. Shor, arXiv:0709.2090 [quant-ph] (2007).

\bibitem{DS08} R. Y. Duan and Y. Y. Shi, Phys. Rev. Lett. \textbf{101}, 020501
(2008).

\bibitem{PAR04} K. R. Parthasarathy, Proc. Indian Acad. Sci. \textbf{114}, 365 (2004).

\bibitem{HLW06} P. Hayden, D. Leung, and A. Winter, Comm. Math. Phys. \textbf{265}, 95
(2006).

\bibitem{DFJY07} R. Y. Duan, Y. Feng, Z. F. Ji, and M. S. Ying, Phys. Rev. Lett. \textbf{98},
230502 (2007).

\bibitem{CMW07} T. S. Cubitt, A. Montanaro, and A. Winter, arXiv: 0706.0705 [quant-ph] (2007).

\bibitem{DXY07} R. Y. Duan, Y. Xin, and M. S. Ying, arXiv: 0708.3559 [quant-ph] (2007).

\bibitem{WS07} J. Walgate and A. J. Scott, arXiv: 0709.4238 [quant-ph] (2007).

\bibitem{CHL+08} T. Cubitt, A. Harrow, D. Leung, A. Montanaro, and
A. Winter, Comm. Math. Phys. \textbf{284}(1), 281 (2008).

\bibitem{SY08} G. Smith and J. Yard, Science \textbf{321}(5897),
1812 (2008).

\bibitem{HAS09} M. B. Hastings, Nature Physics \textbf{5}, 255 (2009).

\bibitem{BDSS06} C. H. Bennett, I. Devetak, P. W. Shor, and J. A. Smolin, Phys.
Rev. Lett. \textbf{96}, 150502 (2006).

\bibitem{LWZG09} K. Li, A. Winter, X. Zou, and G. Guo, arXiv:0903.4308.

\bibitem{SS09} G. Smith and J. A. Smolin, arXiv: 0904.4050.

\bibitem{DMS+00} D. P. DiVincenzo, T. Mor, P. W. Shor, J. A. Smolin,
and B.M. Terhal, Comm. Math. Phys. \textbf{238}, 379 (2003).

\bibitem{BDS97} Charles H. Bennett, David P. DiVincenzo, John A.
Smolin, Phys. Rev. Lett. \textbf{78}, 3217 (1997).

\bibitem{FW07} M. Fukuda and M. M. Wolf, J. Math. Phys. \textbf{48}, 072101
(2007).

\bibitem{NIE99} M. A. Nielsen, Phys. Rev. Lett., \textbf{83}, 436 (1999)

\bibitem{DCX09} R. Y. Duan, J. X. Chen, and Y. Xin, Unambiguous
and zero-error classical capacity of noisy quantum channels,
manuscript in preparation, 2009.

\bibitem{CCH09} T. Cubitt, J. X. Chen, and A. Harrow, arXiv: 0906.2547.
\end{thebibliography}
\end{document}